\documentclass{PoS}
\usepackage{citesort}
\newcommand{\eq}{\begin{eqnarray}}
\newcommand{\en}{\end{eqnarray}}

\title{Three particles in a finite volume}

\ShortTitle{Three particles in a finite volume}

\author{\speaker{Akaki Rusetsky}\\
        Helmholtz-Institut f\"ur Strahlen- und Kernphysik (Theorie) 
and Bethe Center for Theoretical Physics, Universit\"at Bonn, D-53115 Bonn, Germany\\
        E-mail: \email{rusetsky@itkp.uni-bonn.de}}

\abstract{
The volume-dependence of a shallow three-particle bound state in the cubic
box with a size $L$ is studied. It is shown that, in the unitary limit, the energy-level shift
from the infinite-volume position is given by
$\Delta E=c (\kappa^2/m)\,(\kappa L)^{-3/2}|A|^2
\exp(-2\kappa L/\sqrt{3})$, 
where $\kappa$ is the bound-state momentum and 
 $|A|^2$ denotes the three-body analog
of the asymptotic normalization constant, which encodes the information about
the short-range
interactions in the three-body system.
}

\FullConference{The 33rd International Symposium on Lattice Field Theory\\
		 14 -18 July  2015\\
		 Kobe International Conference Center, Kobe, Japan}

\begin{document}

\section{Introduction}

Despite the fact that there have been several attempts to formulate the
L\"uscher-type formalism in case of three 
particles~\cite{Polejaeva,Sharpe1,Sharpe2,BD,DeanLee}, 
the problem is still far from its solution. Conceptually, the central question
here is, whether the finite-volume spectrum  is determined
solely by the {\em observables} in the system of three particles in the infinite
volume. This statement was proved first in Ref.~\cite{Polejaeva} and
subsequently confirmed\footnote{The authors of
 Ref.~\cite{kreuzer} have arrived at the same conclusion on the basis 
of a direct numerical solution of the three-particle equations in a 
finite volume.} in Refs.~\cite{BD,Sharpe1,Sharpe2}. However, the three-particle
quantization condition, given in Refs.~\cite{Sharpe1,Sharpe2}, has still a very 
complicated form. Hence, even though 
the problem is solved in principle, a practical
use of this solution can still require a major effort. On the other hand,
such a formalism is urgently needed to analyze already existing data on the
resonances that may decay into the three-particle final states, e.g., the
Roper resonance~\cite{Roper} or the $a_1(1260)$ meson~\cite{a1}, 
as well as to correctly interpret the
data on the nuclei obtained from lattice QCD and nuclear lattice 
simulations~\cite{NPLQCD,Yamazaki,nuclear-EFT}. 
Since a large amount of far more precise lattice data
is expected to appear in the near future, an approach that can be handled
 with a relative ease, is urgently needed.

The aim of the present work is to study the energy spectrum of the shallow
bound states of the three identical non-relativistic bosons in a finite volume.
Additionally, it is assumed that the two-body scattering length $a$ is 
very large, i.e., one is working in the so-called unitary limit.
Such a choice is justified by the fact that, for this system, the energy
shift from the infinite-volume value can be evaluated analytically in a rather
straightforward manner. Moreover, such systems can be studied by using
numerical methods (see, e.g.,~\cite{kreuzer}), so one can directly verify
the analytic results. In our opinion, the study of the simple systems is 
interesting, first and foremost, because this gives an insight, how the
general quantization condition for the three-particle sector works in practice.
The information, gained in the study of this simple case, can be further used
to design a framework that can be applied to more complex systems.

In the present paper, we do not treat the three-body bound state by using
the quantization condition derived in~\cite{Sharpe1,Sharpe2}. We rather 
follow the original derivation in the position space, which can be found 
in Ref.~\cite{luescher-stable} (see also~\cite{DeanLee,koenig}, where 
the formula for the energy shift is generalized 
to the case of an arbitrary angular momentum). 
The advantage of such a choice is that the role of the 
hyperspherical coordinates becomes very transparent in the position space. 
Namely, we shall see that the asymptotic form of the bound-state wave function
can be most easily written down in terms of the hyperspherical coordinates
that enables one to derive an analytic expression for the energy shift 
in a straightforward way. Needless to say that the same result should be obtainable on the basis of the three-particle quantization 
condition~\cite{Sharpe1,Sharpe2}. Finally, note that the main results of 
the present work are contained in Ref.~\cite{Rios}.

\section{Derivation of the formula}

The Schr\"odinger equation for three identical particles with the mass $m$
in the infinite volume reads as
\eq
\biggl\{\sum_{i=1}^3\biggl(-\frac{1}{2m}\,\nabla_i^2+V({\bf x}_i)\biggr)
+E_T\biggr\}
\psi({\bf r}_1,{\bf r}_2,{\bf r}_3)=0\, ,
\en
where  $\nabla_i=\partial/\partial {\bf r}_i$ , $E_T=\kappa^2/m$, and 
$\kappa$ stands for the three-body bound-state momentum. We further define
the Jacobi coordinates
\eq\label{eq:S-inf}
{\bf x}_i={\bf r}_j-{\bf r}_k\, ,\quad\quad
{\bf y}_i=\frac{1}{\sqrt{3}}\,({\bf r}_j+{\bf r}_k-2{\bf r}_i)\, ,
\en
with $(ijk)=(123),(312),(231)$\footnote{In Eq.~(\ref{eq:S-inf}) no
three-body force is present. Its inclusion, however, proceeds in a 
straightforward manner and does not lead to any complication. This statement
is especially important, if the problem is treated within the field-theoretical framework. It is well known that the model with pair interactions only is not
renormalizable. Therefore, it is perfectly legitimate to ask, what would happen with our finite-volume results in the presence of the three-body force, which is needed in the infinite volume for canceling the ultraviolet divergences.}.

In a finite volume, the potential $V$ is replaced by
\eq
V_L({\bf x}_i)=\sum_{{\bf n}\in\mathbb{Z}^3}V({\bf x}_i+{\bf n}L)\, ,
\en
and the Schr\"odinger equation takes the form
\eq
\biggl\{\sum_{i=1}^3\biggl(-\frac{1}{2m}\,\nabla_i^2+V_L({\bf x}_i)\biggr)
+E_L\biggr\}
\psi_L({\bf r}_1,{\bf r}_2,{\bf r}_3)=0\, .
\en
Next, let us choose the CM frame. Due to the translational invariance,
the bound-state wave functions depend on two
Jacobi coordinates ${\bf x}_i,{\bf y}_i$. 
The Bose symmetry in addition implies 
$\psi({\bf x}_i,{\bf y}_i)=\psi({\bf x}_k,{\bf y}_k)\, ,~i,k=1,2,3,$. 
The wave function $\psi_L$ obeys the same relations.

Our aim here is to evaluate the finite-volume shift $\Delta E=E_T-E_L$.
To this end, we define the trial wave function 
\eq
\psi_0
=\sum_{{\bf n},{\bf m}}
\psi\biggl({\bf x}_1-({\bf n}+{\bf m})L,{\bf y}_1+\frac{1}{\sqrt{3}}\,
({\bf n}-{\bf m})L\biggr)\, .
\en
Introducing the notation $H_L=\sum_{i=1}^3\biggl(-\frac{1}{2m}\,\nabla_i^2+V_L({\bf x}_i)\biggr)$, it can be checked that $\psi_0$ obeys the equation
$(H_L+E_T)\psi_0=\eta$, where
\eq\label{eq:eta}
\eta=\sum_{{\bf n},{\bf m}}\hat V_{\bf nm}
\psi\biggl({\bf x}_1-({\bf n}+{\bf m})L,{\bf y}_1+\frac{1}{\sqrt{3}}\,
({\bf n}-{\bf m})L\biggr)
\en
and
\eq
\hat V_{\bf nm}=
\sum_{{\bf k}\neq-{\bf n}-{\bf m}}V({\bf x}_1+{\bf k}L)
+\sum_{{\bf k}\neq {\bf n}}V({\bf x}_2+{\bf k}L)
+\sum_{{\bf k}\neq {\bf m}}V({\bf x}_3+{\bf k}L)\, .
\en
Since the potential $V({\bf x})$ is short-ranged,
$\eta\propto \exp(-{\rm const}\cdot \kappa L)$.

It can be
 verified that the energy level shift to all orders in perturbation theory
 is given by 
\eq
\Delta E=\frac{\langle\psi_0|T|\psi_0\rangle}{\langle\psi_0|\psi_0\rangle}\, ,\quad\quad 
T=(H_L+E_T)-(H_L+E_T)QGQ(H_L+E_T)\, ,
\en
where
\eq
G=\frac{1}{H_L+E_L}\, ,\quad\quad
Q=\frac{|\psi_0\rangle\langle\psi_0|}{\langle\psi_0|\psi_0\rangle}\, .
\en
The leading exponential correction to the energy shift is given by
\eq
\Delta E=\frac{\langle\eta|\psi_0\rangle}{\langle\psi_0|\psi_0\rangle}
+\cdots\, .
\en
Here, the ellipses stand for higher-order terms.  

What remains to be done is to substitute the expression for $\eta$ from 
Eq.~(\ref{eq:eta}) into the above expression and to single out the
leading exponential contribution at large $L$. Since $V({\bf x})$ is a 
short-range potential, it can be readily seen that at least
one of the wave functions in the overlap integral should be evaluated at
a large value of the hyperradius 
$R=\frac{1}{\sqrt{2}}\,({\bf x}_1^2+{\bf y}_1^2)^{1/2}$. Since the wave functions
decrease exponentially with $R$, the leading exponential
correction can be obtained by minimizing the sum of two hyperradii
in the wave function. Considering all possible terms in Eq.~(\ref{eq:eta}),
one may conclude that the leading contribution is defined by
\eq\label{eq:master}
\Delta E=6\cdot 2\cdot 3\int d^3{\bf x}_1d^3{\bf y}_1 
\psi({\bf x}_1,{\bf y}_1)V({\bf x}_1)
\psi\biggl({\bf x}_1-{\bf e}L,{\bf y}_1+\frac{1}{\sqrt{3}}\,{\bf e}L\biggr)+\cdots\, ,
\en
where ${\bf e}=(0,0,1)$ denotes a unit vector and the ellipses stand for the
exponentially 
suppressed terms. In this formula, the infinite-volume wave 
function $\psi$ is normalized to unity. The factor in front of the integral 
reflects the symmetries: 6 for different orientations of the unit vector 
${\bf e}$, 2 for different signs in the second argument of the wave function
${\bf y}_1\pm \frac{1}{\sqrt{3}}\,{\bf e}L$, and 3 for three different 
pair potentials.

\section{The energy shift}

The leading-order energy shift is given by Eq.~(\ref{eq:master}). As 
mentioned above, in the calculation of the overlap integral it suffices
to replace one of the wave functions by its asymptotic tail, which does not
depend on the short-range details of the interaction. The wave function takes
a universal form in the unitary limit, which in the context of our
problem means that the two-body scattering length $a\geq L$.
On the other hand, it is assumed that the interaction range vanishes.
Under these assumptions, in the 
configuration space, the wave function can be approximated by the
well-known universal expression (see, e.g.~\cite{BraatenHammer})
\eq\label{eq:psi}
\psi({\bf x}_1,{\bf y}_1)=A{\cal N}R^{-5/2}f_0(R)\sum_{i=1}^3
\frac{\sinh(s_0(\pi/2-\alpha_i))}{\sin(2\alpha_i)}
\doteq\sum_{i=1}^3\phi(R,\alpha_i)\, ,
\en
where
\eq
f_0(R)=R^{1/2}K_{is_0}(\sqrt{2}\kappa R)
\en
and $K_\nu(z)$ denotes the Bessel function. Here, 
$\alpha_i=\arctan(|{\bf x}_i|/|{\bf y}_i|)$  
and the numerical constant $s_0\simeq 1.00624$.

Note that the above expression is valid {\em almost} everywhere in the 
configuration space. The exception is given by the configurations, when two
out of three particles stay close together as the hyperradius grows to 
infinity. We are interested, however, in the overlap integral over all 
configurations. Since the exceptional configurations have zero measure in the limit of the vanishing effective range, 
they will
not give contribution to the final result, provided the rest of the integrand
is not singular there. This qualitative discussion gives a clear hint, why the 
finite-volume  spectrum is, at the end, determined solely by the infinite-volume
observables, even in the presence of the so-called ``spectator'' diagrams.

Further, the normalization coefficient ${\cal N}$ in Eq.~(\ref{eq:psi})
is chosen so that 
\eq
\int d^3{\bf x}_1 d^3{\bf y}_1 |\psi({\bf x}_1,{\bf y}_1)|^2=|A|^2\, .
\en
The quantity $|A|^2$ is a very important characteristic of a bound state since
it encodes the information about the short-range dynamics in the system.
It could be interpreted as a three-body analog of
the asymptotic normalization coefficient for the wave function.
Consider, for example, the situation, when the long-range effects dominate
and the true wave function coincides with the asymptotic wave function almost
everywhere. Since the true wave function is normalized to unity, the quantity
$|A|^2$ should be very close to one. In the opposite case, when almost the whole
wave function is concentrated at small distances, the quantity $|A|^2$ should
be very small. It is seen that measuring the quantity $|A|^2$ on the lattice 
will allow one to judge about the nature of a bound state, in analogy to the
two-body case (see, e.g.,~\cite{GuoRios} and references therein).

After the general discussion, we turn to the evaluation of the
overlap integral in Eq.~(\ref{eq:master}) by using
the explicit wave function from Eq.~(\ref{eq:psi}). 
The second wave function in the integral can be written as
\eq
\psi\biggl({\bf x}_1-{\bf e}L,{\bf y}_1+\frac{1}{\sqrt{3}}\,{\bf e}L\biggr)
=\sum_{i=1}^3\phi(R',\alpha'_i)\, .
\en
As $L\to\infty$, the quantity $R'$ becomes
\eq
R'=\frac{(({\bf x}_1-{\bf e}L)^2+
({\bf y}_1+{\bf e}L/\sqrt{3})^2)^{1/2}}{\sqrt{2}}\to \sqrt{\frac{2}{3}}\,L
-\frac{\sqrt{3}}{2\sqrt{2}}\,{\bf e}\cdot{\bf x}_1+\frac{1}{2\sqrt{2}}{\bf e}\cdot{\bf y}_1+\cdots\, ,
\en
whereas the angular variables tend to the following limiting values:
\eq
\tan\alpha_1'&\!\!=\!\!&\frac{|{\bf x}_1-{\bf e}L|}{|{\bf y}_1+{\bf e}L/\sqrt{3}|}\to\sqrt{3}+\cdots\, ,
\nonumber\\
\tan\alpha_2'&\!\!=\!\!&\frac{|{\bf x}_2+{\bf e}L|}{|{\bf y}_2+{\bf e}L/\sqrt{3}|}\to\sqrt{3}+\cdots\, ,
\nonumber\\
\tan\alpha_3'&\!\!=\!\!&\frac{|{\bf x}_3|}{|{\bf y}_3-{2\bf e}L/\sqrt{3}|}\to
\frac{\sqrt{3}}{2}\,\frac{|{\bf x}_3|}{L}+\cdots
\nonumber\\
&\!\!=\!\!&\frac{\sqrt{6}R\sin\alpha_3}{2L}+\cdots\, .
\en
The expansion of the angular part of the second wave function yields
\eq
\sum_{i=1}^3
\frac{\sinh(s_0(\pi/2-\alpha'_i))}{\sin(2\alpha'_i)}\to
\frac{L}{\sqrt{6}R}\,\frac{\sinh(\pi s_0/2)}{\sin(\alpha_3)}+\cdots\, .
\en
Next, using the equation
\eq
\psi({\bf x}_1,{\bf y}_1)V({\bf x}_1)=\biggl(\frac{1}{m}\,\biggl(
\frac{\partial^2}{\partial {\bf x}_i^2}+\frac{\partial^2}{\partial {\bf y}_i^2}
\biggr)-E_T\biggr)\phi(R,\alpha_1)
\en
and the asymptotic expression for the hyperradial wave function
\eq
f_0\biggl(\sqrt{\frac{2}{3}}L\biggr)\to\sqrt{\frac{\pi}{2}}\exp\biggl(-\frac{2\kappa L}{\sqrt{3}}\biggr)
\exp\biggl(\frac{\sqrt{3}\kappa}{2}{\bf e}\cdot{\bf x}_1-\frac{\kappa}{2}\,{\bf e}\cdot{\bf y}_1\biggr)
\frac{1}{(\sqrt{2}\kappa)^{1/2}}+\cdots\, ,
\en
one may evaluate the overlap intehral in Eq.~(\ref{eq:master}). The final result
looks as follows:
\eq\label{eq:final}
\Delta E=c (\kappa^2/m)\,(\kappa L)^{-3/2}|A|^2
\exp(-2\kappa L/\sqrt{3})+\cdots\, ,
\en
where
\eq
c\simeq -96.351
\en
is a numerical constant expressed in terms of $s_0$,
and the ellipses stand for the sub-leading terms\footnote{In difference to the two-particle case, here the sub-leading terms which are down only by {\em powers} of $L$, are also present.} in $L$.

The Eq.~(\ref{eq:final}) displays our main result.
Measuring the binding energy at different volumes, one may determine 
the infinite-volume quantities $E_T$  and $|A|^2$ 
through the extrapolation procedure.

\section{Conclusions}

The equation~(\ref{eq:final}) is an explicit
prediction of the volume dependence for a genuine three-body observable.
Understanding this result on the basis of the three-body quantization 
condition~\cite{Sharpe1,Sharpe2} would enable one to gain more insight
into the structure of three-body scattering equations in a finite volume and
would therefore facilitate the formulation of a framework that can be 
conveniently used to analyze the lattice data on the three-particle systems.

In the future, we plan to move, step by step, away from the approximations,
which were used to derive the result given in Eq.~(\ref{eq:final}). Namely,
we would like to study the effects of a finite scattering length, perform 
an effective-range expansion in the potential and include the 
partial-wave mixing.

\begin{acknowledgments}
The authors would like to thank S.~Bour, M.~D\"oring, E.~Epelbaum, H.-W.~Hammer, M.~Hansen, 
M.~Jansen, D.~Lee and S.~Sharpe for useful discussions.
This work is partly supported by the EU
Integrated Infrastructure Initiative HadronPhysics3 Project  under Grant
Agreement no. 283286. We also acknowledge the support by the DFG (CRC 16,
``Subnuclear Structure of Matter'' and CRC 110, ``Symmetries and the Emergence of Structure in QCD'')
This research is supported in part by Volkswagenstiftung
under contract no. 86260.
\end{acknowledgments}

\end{document}